\documentclass[twocolumn,showpacs,amssymb,nobibnotes,superscriptaddress,aps,pra]{revtex4}

\usepackage{enumerate,amsmath}
\usepackage{graphicx}
\usepackage{times}
\begin{document}

\title{Computational Power
and Correlation in Quantum Computational Tensor Network}

\author{Keisuke Fujii}
\affiliation{
Graduate School of Engineering Science, Osaka University,
Toyonaka, Osaka 560-8531, Japan}

\author{Tomoyuki Morimae}
           \affiliation{
 Universit\'e Paris-Est Marne-la-Vall\'ee, 77454 Marne-la-Vall\'ee
 Cedex 2, France
}
 \affiliation{
Interactive Research Center of Science (IRCS), Tokyo Institute of Technology,
2-12-1 Ookayama, Meguro-ku, Tokyo 152-8550, Japan
}
\affiliation{
Controlled Quantum Dynamics Theory Group, Imperial College London,
London SW7 2AZ, United Kingdom
}
\date{\today}
\begin{abstract}
We investigate relationships between computational power
and correlation in resource states for quantum computational tensor network, 
which is a general framework for measurement-based quantum computation.
We find that if the size of resource states is finite, 
not all resource states allow correct projective measurements in the correlation space,
which is related to non-vanishing two-point correlations in the resource states.
On the other hand, for infinite-size resource states,
we can always implement correct projective measurements 
if the resource state can simulate arbitrary single-qubit rotations,
since such a resource state exhibits exponentially decaying two-point correlations.
This implies that a many-body state whose
two-point correlation cannot be upperbounded by an exponentially decaying
function cannot simulate arbitrary single-qubit rotations.
\end{abstract}


\maketitle

\section{Introduction}
Quantum many-body states, which have long been the central research
objects in condensed matter physics and statistical physics, 
are now attracting the
renewed interest in quantum information science.
In particular, in measurement-based quantum computation (MBQC)~\cite{OWC1,OWC2},
the role of quantum many-body states played in quantum computation
is very clear:
once the special resource state, which is called the
cluster state, is prepared, universal quantum computation is
possible with adaptive local measurements on each qubit.
This clear separation between the 
resource preparation and the execution of computation itself
has also relaxed
the requirements for experimental realization of quantum computation~\cite{
Nielsen04,Browne05,Duan05,Barrett05,3D1,3D2,3D3,FT10,FM12}.
Recently, the concept of quantum computational tensor network (QCTN),
which is a novel framework of MBQC
on general many-body states, has been proposed~\cite{Gross07_1,Gross07_2,Gross07_3}.
The most innovative feature of
this framework is that the resource state is represented in
the matrix-product (or tensor-product) form~\cite{Fannes91,Verstraete08,Cirac09} 
and universal quantum computation is simulated in
the virtual linear space, which is called the correlation space (CS)~\cite{Gross07_1,Gross07_2,Gross07_3},
where matrices (or tensors) live.

How does the behavior of a physical quantity 
in a many-body resource state affect the computational
power of a quantum computer?
If one wants to construct a bridge between quantum many-body physics
and quantum information science,
one cannot avoid tackling such a question.
For example, in Refs.~\cite{OWCent1,OWCent2}, the beautiful result was obtained
that if a many-body resource 
state has too much entanglement, it
is useless for MBQC.
In Ref.~\cite{Nestent}, a criterion of the amount of entanglement
for universal preparators in graph states was obtained.
In Ref.~\cite{Morimaeent}, a relationship between the amount of entanglement
and the gate fidelity in the one-way model was derived.
These results have clarified important relationships between
entanglement in many-body states
and the computational power of MBQC.

Surprisingly, less attention has been paid to
relationships between
more traditional physical quantities,
namely, a two-point correlation and the computational power
of a quantum computer.
Obviously, a two-point correlation is one of the most
important research
subjects in traditional quantum many-body physics,
such as condensed matter physics and statistical physics.
If we consider applications of quantum many-body states
in condensed matter physics
to quantum computation \cite{Verstraete04}, it is natural to ask
how the behavior of a two-point correlation affects the computational
power of a quantum computer.
It has long been known in quantum information science that
two-point correlations exactly vanish within a finite
distance in the cluster state.
Such a correlation property prohibits the cluster state 
from being the exact ground state of any physical Hamiltonian~\cite{Nielsen06,Nest08,Chen11}.
Recently, it was shown that the ground state of the 
Affleck, Kennedy, Lieb, and Tasaki (AKLT) model~\cite{AKLT}, which exhibits the
exponentially decaying two-point
correlation, enables universal MBQC~\cite{Brennen08,Gross07_1,Gross07_2,Gross07_3}.
This result suggests that we might be able to further explore other resource
states in the pool of quantum many-body states including
ones which exhibit
polynomially decaying or non decaying two-point correlations.
Considering the fact that there are plenty of many-body states
which exhibit such a long-range two-point correlation
in condensed matter physics, it is very interesting to study
whether we can use long-range-correlated many-body states as
resources of MBQC.

In this paper, 
we investigate relationships between
computational power and two-point correlations in QCTN
considering both finite and infinite-size resource states.
Specifically, we here address under what conditions
we can simulate projective measurements 
and arbitrary single-qubit rotations in QCTN,
both of which are important properties of a universal resource for 
MBQC~\cite{predict}.

We find that projective measurements are not simulated correctly 
in QCTN on certain resource states of finite size,
such as the AKLT states with a specific boundary condition.
We therefore introduce a class of finite-size matrix-product states (MPSs), 
namely, {\it normalizable}
resources, which allow correct projective measurements in the CSs.
Since two-point correlations exactly vanish within a finite distance
in normalizable resources, they are considered as
cluster like resources.
In order to perform correct projective measurements
on finite resource states with non vanishing two-point
correlations, we introduce
another class of MPSs.
We also mention an alternative way of simulating projective measurements
indirectly by using the downloading method~\cite{Cai09}. 

We further show that if a resource state
can simulate arbitrary single-qubit rotations,
it can also perform correct projective measurements 
in the infinite-size limit, that is, it is {\it asymptotically normalizable}.
Accordingly, normalizability (at least asymptotic sense)
is a necessary and sufficient condition
for a universal quantum computational wire,
where both correct projective measurements
and arbitrary single-qubit rotations can be simulated.
This result is obtained by showing a theorem
about convergence of a stochastic unitary map.
Furthermore, this theorem also tells us 
that a many-body state whose two-point correlation
cannot be upperbounded by an exponentially decaying function, 
such as the Greenberger-Horne-Zeilinger (GHZ),
the Ising antiferromagnetic, and the $W$ states,
cannot be used to perform an arbitrary single-qubit rotation.
These findings highlight interesting 
relationships between computational power
and correlation in many-body entangled states.

This paper is organized as follows.
In Sec. I, we briefly review QCTN \cite{Gross07_1,Gross07_2,Gross07_3},
which is the most general framework for MBQC.
In Sec. II, we investigate simulations of projective measurements 
in the QCTN framework and specify under what situation
one can accurately simulate projective measurements
in the resource states of finite size.
In Sec. III,
we extend the notion of normalizability
to the resource states of infinite size, namely asymptotic normalizability,
and address a fundamental relationship between
asymptotic normalizability and computational power 
by providing a theorem which makes a bridge between the ability to 
perform an arbitrary single-qubit rotation
and behavior of the two-point correlations.
Section IV is devoted to the conclusion.

\section{Quantum computational tensor network}
\begin{figure}
\centering
\includegraphics[width=85mm]{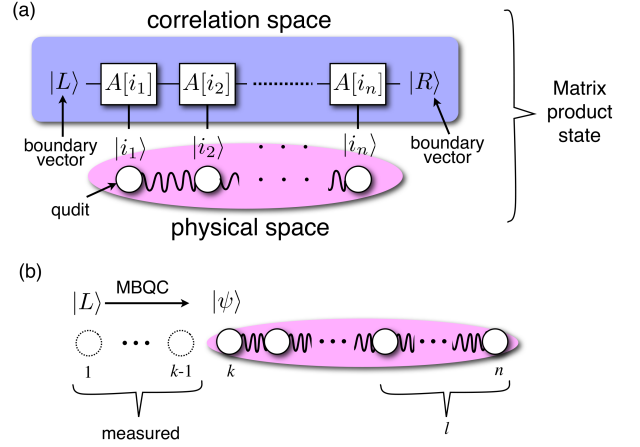}
\caption{(a) Matrix product state. (b) By measuring from the 1st
to $(k-1)$th qudits, a single qubit rotation is simulated
in the correlation space, $|L\rangle \rightarrow |\psi\rangle$.
In the normalizable resource with a finite $l$,
projective measurements are simulated correctly
at every position which is more than $l$ qudits away 
from the right boundary.}
\label{fig1}
\end{figure}

Let us consider an MPS \cite{Fannes91,Verstraete08}:
\begin{eqnarray*}
|\Psi (R,L)\rangle _{1}^{n} \equiv  \sum _{ i _1,\cdots, i_n} \langle R| A[i_{n}] \cdots A[i_{1}]|L\rangle | i _{n}  \cdots i_{1}\rangle,
\end{eqnarray*}
where $A[i_{k}]$ and $|R\rangle,|L\rangle$ ($\langle R|R\rangle$=$\langle L |L\rangle =1$) are a $D\times D$ matrix
and $D$-dimensional boundary vectors \cite{Gross07_1,Gross07_2,Gross07_3}, respectively,
which live in the virtual Hilbert space, the so-called {\it correlation space} (CS) \cite{Gross07_1,Gross07_2,Gross07_3} [see Fig. \ref{fig1} (a)].
The $|i_{k}\rangle$ ($i_k = 0, \cdots  d-1$ and $\langle i |j \rangle = \delta _{ij}$) denotes the $k$th physical qudit.
Since we are interested in the simulation of unitary operations
in the CS, 
we assume that the MPS matrices $A[i]$ of a resource state
can be described as $A^{\dag}[i] A[i]=c_i I$ with positive
real numbers $c_i$ by choosing
the physical basis properly,
where $\sum _{i=0}^{d-1} c_i=1$ (i.e. $\sum _{i=0}^{d-1} A^{\dag}[i]A[i]=I$).
We here consider the case of $D=2$, that is,
the CS simulates a single qubit.
We can simulate a quantum gate in the CS
by performing a projective measurement in the basis $\{ | \alpha _{j} \rangle \equiv \sum _{i=0}^{d-1} \alpha _{ji} |i\rangle  \}$ 
($i,j = 0, \cdots , d-1$ and $\langle \alpha _i | \alpha _j \rangle =\delta _{ij}$) 
on a physical qudit.
The postmeasurement state with the measurement outcome $\alpha _{j}$ reads
\begin{eqnarray}
&&\sum _{ i _2,  \cdots, i_n} \langle R| A[i_{n}] \cdots A[i_2]  A[ \alpha _j ]  |L\rangle
| i_{n} \cdots i_2 \rangle \otimes | \alpha_j \rangle
\nonumber \\
&=& 
|\Psi (R,L')\rangle _2^n  \otimes  \sqrt{a_j} | \alpha _j \rangle,
\label{eq1}
\end{eqnarray}
where $A[\alpha _j] \equiv \sum _{i=0}^{d-1} \alpha^{*} _{ji} A[i]$ and   
 $|L'\rangle \equiv A[\alpha _j]/\sqrt{a_j} |L\rangle$ with $a_j I  \equiv 
 A^{\dag}[\alpha _j ]A [\alpha _j] $.
The postmeasurement state can be 
understood that the boundary vector $|L\rangle$
is transformed to $|L'\rangle$ 
by the single-qubit operation $A[\alpha _j]/\sqrt{a_j}$.
So far, the normalization of an MPS has not been considered properly. 
In order to address this point,
we define the normalization factor as
\begin{eqnarray*}
f_n(|R\rangle , |L\rangle) \equiv 
\langle R | (\mathcal{A} ^{\circ n} |L\rangle \langle L| ) |R\rangle ,
\end{eqnarray*}
where $\mathcal{A} \rho = \sum _{i=0}^{d-1} A[i] \rho A^{\dag} [i]$.
We denote the normalized MPS as 
\begin{eqnarray*}
| \bar \Psi (R,L)\rangle _{1}^{n} \equiv |\Psi (R,L)\rangle _{1}^{n}/\sqrt{f_n(|R\rangle , |L\rangle) }.
\end{eqnarray*}

\section{Simulations of projective measurements}
\subsection{Condition for correct projective measurements}
Let us consider a situation where
from the first to $(k-1)$th physical qudits are measured
in order to implement a single-qubit rotation in the CS,
and the boundary vector $|L\rangle$
is transformed to $|\psi\rangle$ [see Fig. \ref{fig1} (b)]:
\begin{eqnarray*}
|\bar\Psi (R,\psi)\rangle _{k}^{n} = \sum _{i_k \cdots i_n} \frac{\langle R| A[i_n] \cdots A[i_k] |\psi\rangle
 |i_n \cdots i_k\rangle}
{ \sqrt{f_{n-k+1}(|R\rangle , |\psi\rangle)}}.
\end{eqnarray*}
Then, we want to perform a projective measurement 
on the state $|\psi \rangle$ in the CS
in a basis $\{ |0\rangle , |1\rangle\}$.
To this end, without loss of generality,
the MPS matrix $A[i]$ can be written
as 
\begin{eqnarray*}
A[0] &=& \sqrt{c_m/2}( | \phi _0 \rangle \langle 0|+| \phi _1 \rangle \langle 1|)
\\
A[1] &=& \sqrt{c_m/2}(| \phi _0 \rangle \langle 0|-| \phi _1 \rangle \langle 1|)
\end{eqnarray*}
by choosing a proper physical basis, 
where $|\phi _0\rangle$ and $|\phi _1\rangle$
are orthonormal states and $c_m$ is a real number.
If we do the measurement $\mathcal{M}=\{ |m_0 \rangle =(|0\rangle + |1\rangle)/\sqrt{2},|m_1 \rangle =(|0\rangle - |1\rangle)/\sqrt{2}, |2\rangle , \cdots, |d-1\rangle\}$ on the $k$th physical qudit
and obtain the outcome $m_j$ ($j=0,1$),
the probability $p_j$ of obtaining 
$m_j$ ($j=0,1$) is calculated to be
\begin{eqnarray*}
p_j= \frac{c_m f_{n-k}(|R\rangle,|\phi _j \rangle)  |\langle j  |\psi \rangle|^2}{f_{n-k+1}(|R\rangle,|\psi\rangle)}.
\end{eqnarray*}
If the above operation properly simulates the projective measurement in the CS,
$p_0/p_1 = |\langle 0 | \psi \rangle |^2/ |\langle 1 | \psi \rangle |^2$ must be satisfied.
This leads to
\begin{eqnarray}
f_{n-k} (|R\rangle ,|\phi _0\rangle) = f_{n-k} (|R\rangle ,|\phi _1\rangle).
\label{cond}
\end{eqnarray}
If a resource state does not satisfy condition (\ref{cond}),
correct projective measurements are not simulated in the CS.

\subsection{Normalizable resources}
In order to perform correct QCTN,
we have to find MPSs which satisfy condition (\ref{cond}).
Let us first consider a sufficient condition of Eq. (\ref{cond}).
We call a resource state {\it normalizable},
if there exists a natural number $l$ ($l \leq n$) such that 
\begin{eqnarray*}
f_{n-k} (|R\rangle,| \psi \rangle)=1/2,
\end{eqnarray*}
for all $|\psi\rangle$, $|R\rangle$ and all $k \leq n-l$.
Obviously normalizable resources satisfy condition (\ref{cond}) as long as $n-k\geq l$,
which means that we can perform correct projective measurements 
in the CS
at every position which is more than $l$ qudits away 
from the left boundary.
Later we will see that normalizability in the limit of $l \rightarrow \infty$ is also 
a necessary condition for an arbitrary single-qubit rotation.

A resource state is normalizable,
if and only if the map $\mathcal{A}^{\circ l}$ ($ l \le n-k $) is a completely depolarizing channel:	
\begin{eqnarray*}
\mathcal{E} \rho  =1/4  \sum_{i=0}^{3} U \sigma _{i} V \rho V^{\dag} \sigma _{i} U^{\dag},
\end{eqnarray*}
where $\sigma _{i}$ are the Pauli matrices ($\sigma _0=I$),
and $U$, $V$ are arbitrary unitary operators 
(see Appendix A).
If a finite-size resource state ($n < \infty$) is normalizable, 
$l$ is also finite. 
Let us first consider the case of $l=1$.
In this case, $\mathcal{A}=\mathcal{E}$,
which means that the dimension of a physical system is at least $d=4$,
and in such a case the MPS matrices are given by $A[i]= U\sigma _iV /2$.
It is obvious that such a resource state 
with properly chosen $U$ and $V$ allows
an arbitrary single-qubit rotation.
Next we consider the case of $l=2$.
The minimum dimension of a qudit which satisfies ${\mathcal{A}}^{\circ 2} =\mathcal{E}$
is $d=2$, where the MPS matrices are given, for example, by
$A[i] A[j] = U \sigma _{i+2j} V$ ($i,j =0,1$). 
Specifically, if we choose $U$ and $V$ in such a way that
$U\sigma _1U^{\dag}=\sigma _1$, $U \sigma _3 U^{\dag}= -i \sigma _2$, 
$U \sigma _2 U^{\dag}=\sigma _3$ and $V=U^{\dag}$, 
the MPS matrices become $A[0]=H/\sqrt{2}$ and $A[1]=HZ/\sqrt{2}$,
which is the matrices for the cluster state \cite{OWC1,OWC2,Gross07_1,Gross07_2,Gross07_3}.
Since the above two- and four-dimensional resource states are
normalizable with a finite $l$, we can simulate correct projective measurements
in the CSs even if their size is finite.

The above definition of normalizability also has an important 
physical meaning: if a resource state is normalizable with a finite $l$ 
(i.e., $\mathcal{A}^{\circ l} =\mathcal{E}$),
any two-point correlation
$\langle O^a _k O^b _{k+r} \rangle _{\Psi} - \langle O^a _k \rangle _{\Psi} \langle O^b _{k+r} \rangle _{\Psi} $
vanishes for all $r >l$ 
(see Appendix B),
where 
$O_k^{a,b}$ is an arbitrary operator on the $k$th qudit,
and
$\langle O \rangle _{\Psi} \equiv \langle  \bar \Psi (L,R) | O  | \bar \Psi (L,R) \rangle _1^n$.
Thus the possibility of projective measurements in QCTN
with finite-size resource states 
is related to their vanishing two-point correlations.
In the sense that two-point correlations exactly vanish with a finite distance,
the normalizable resources with a finite $l$
can be viewed as cluster like resource states \cite{OWC1,OWC2}.
Then, it is natural to ask,
can we simulate correct projective measurements on
the resource states with non vanishing two-point correlations?

\subsection{A class for finite QCTN}
In order to answer that question,
we here find a class of finite-size MPSs,
which is not necessarily normalizable with a finite $l$, 
but can satisfy condition (\ref{cond}) with a finite $l$
by properly choosing the boundary condition $|R\rangle$.
This class of MPSs includes
a wide variety of resource states,
such as the AKLT states with specific boundary conditions \cite{Gross07_1,Gross07_2,Gross07_3,Brennen08,comment}.
As a canonical form of MPSs in the class,
we take the matrices as
\begin{eqnarray}
A[0]&=&\sqrt{c_m/2 }  e^{ i \theta _0  Y}, 
A[1]=\sqrt{c_m/2}  e^{i \theta _0 Y} Z,
\nonumber \\
A[j]&=&\sqrt{c_j} e^{i \theta _j Y} B_j 
\label{eq4}
\end{eqnarray}
for $j=2,\cdots , d-1$,
where $B_j \in \{ I,X,Y,Z \}$ 
and $c_m + \sum _{j=2}^{d-1} c_j =1$, i.e., $\sum _{i=0}^{d-1} A^{\dag}[i]A[i]=I$.
We can easily show that this class of MPSs 
with properly chosen $\theta _i$ is capable of 
performing an arbitrary single-qubit rotation.
Since $A[i]$ consists of the Pauli matrices and rotations about the $Y$ axis,
the map $\mathcal{A}$ transforms the Pauli matrices as follows
\begin{eqnarray*}
\mathcal{A} Y = \gamma _{22} Y,
\mathcal{A} X = \gamma _{11} X+ \gamma _{13} Z,
\mathcal{A} Z = \gamma _{31} X+ \gamma _{33} Z,
\end{eqnarray*}
where $\gamma _{ij}$ are some real numbers.
Using this fact and 
$|\phi _j \rangle \langle \phi _j| = I /2
+ (-1)^{j} ( \cos 2 \theta _0 Z - \sin 2 \theta _0 X )/2$,
we can conclude that 
the normalization factor takes the following form:
\begin{eqnarray*}
f_{n-k} (|R\rangle , |\phi _j \rangle )
=\langle R |( I/2 + g_3^{j} Z + g_1^{j}X ) |R\rangle,
\end{eqnarray*}
with $g_{3,1}^{j}$ being some real numbers,
which depend on $n$ and $k$ in general. 
Then, if we choose $|R\rangle \langle R|= (I \pm Y)/2$,
$f_{n-k} (|R\rangle , |\phi _0 \rangle )=  f_{n-k} (|R\rangle , |\phi _1 \rangle )=1/2$
for all $k \leq n$.
That is, by choosing the boundary vector as $|R\rangle=(I \pm Y)/2$,
we can perform correct projective measurements at every position \cite{comment4}.
Instead of $Y$ and $Z$,
we can also use other combinations of two Pauli operators
in the canonical form (\ref{eq4}).
Specifically, if we set $d=3$, $\theta _0= -\pi / 2$, $\theta _2= 0$, $B_2=Z$, and $c_m/2=c_2=1/3$,
the matrices become $A[0]= XZ/\sqrt{3}$, $A[1]=X/\sqrt{3}$, and 
$A[3]= Z /\sqrt{3}$,
which is the AKLT state \cite{Brennen08}
(we can also show that the AKLT-like state \cite{Gross07_1,Gross07_2,Gross07_3}
is an MPS of this class).

\subsection{Download as an indirect way to perform 
projective measurements}
We here explore another way to simulate projective measurements indirectly.
In Ref. \cite{Cai09},
a method has been proposed 
to download a quantum state from the CS to the physical space.
In their method, we first project the $k$th qudit of $|\bar \Psi (R,\psi)\rangle _k^n$
to the subspace spanned by $\{ |m_0\rangle , |m_1\rangle \}$
(if we fail the projection, we repeat it again until it succeeds).
Next we rotate the state $|\phi _j\rangle $ to $|m_j \rangle$ ($j=0,1$) via MBQC.
Then we now have 
\begin{eqnarray*}
\sum _{j=0,1} | \bar \Psi (R,m _j)\rangle _{k'}^{n}
 \otimes \sqrt{\frac{c_u c_m f_{n-k'+1}(|R\rangle , |m _j \rangle)}{f_{n-k+1}(|R\rangle , |\psi\rangle)}}\langle j | \psi \rangle  |m_j\rangle 
\end{eqnarray*}
where $k' \geq k+1$, and the factor $c_u$ comes from 
the measurements for the rotation $|\phi _j\rangle \rightarrow |m_j \rangle$ in the CS.
Finally we perform the projective measurement $\mathcal{M}$
on the $k'$th qudit \cite{comment2}. 
The postmeasurement state corresponding to the measurement 
outcome $m_l$ ($l=0,1$) is given by
\begin{eqnarray*}
&& \sum _{j=0,1} | \bar \Psi (R,\phi _{l})\rangle _{k'+1}^{n} \otimes  |m_l\rangle 
\\
&& \otimes  HZ^{l}
\sqrt{\frac{c_u c_m^2 f_{n-k'}(|R\rangle , |\phi _l \rangle)}{2f_{n-k+1}(|R\rangle , |\psi\rangle)}}
 \langle j | \psi \rangle  |j\rangle .
\end{eqnarray*}
The coefficient of  $\langle j | \psi \rangle  |j\rangle$ does not depend
on $j$, and hence we can download the state from 
the CS to the physical space,
even if the resource state is of finite size.
This means that MBQC on finite-size resource states
is universal state preparator \cite{Nestent}.
Surprisingly, the above fact means that for finite-size resource states
universal state preparator is more ubiquitous property than the simulator of a classical output.
Of course we can perform projective measurements indirectly on a 
quantum state in the CS by downloading it to the physical space and
measuring the downloaded qudit. 
However, as discussed in Ref. \cite{Cai09},
a faithful downloading process takes 
a time step in proportion to the correlation length
of the resource state,
which results in additional operations.
Since such additional operations
can cause additional decoherence effects,
it would be better to avoid doing download if we need only classical outputs.

\section{Asymptotic Normalizability and Computational power}
Next we investigate a relationship between normalizability
and possibility of an arbitrary single-qubit rotation
by extending the notion of normalizability to the infinite-size resource states.
We call a resource state {\it asymptotically 
normalizable} if it is normalizable only in the limit $l \rightarrow \infty$.
For example, in the case of the AKLT state \cite{Brennen08},
the matrices are given by $A[0]=X/\sqrt{3}$, $A[1]=XZ/\sqrt{3}$, and $A[2]=Z/\sqrt{3}$,
and the normalization factor can be calculated to be
\begin{eqnarray*}
f_{n-k} ( |R\rangle , |\psi \rangle )
= 1/2 + (-1/3)^{n-k} \sum _{i=1}^{3} e_i \langle R | \sigma _i | R\rangle/2,
\end{eqnarray*}
where $|\psi \rangle \langle \psi | = I/2+\sum _{i=1}^{3} e_i \sigma _i/2$.
In the limit $l \rightarrow \infty$,
$f_{n-k}( |R\rangle , |\psi \rangle ) \rightarrow 1/2$,
and hence the AKLT state is asymptotically normalizable.

To make a bridge between asymptotic normalizability and 
possibility of an arbitrary single-qubit rotation, 
in the following, we show a theorem about convergence of 
completely-positive-trace-preserving (CPTP) maps,
which allows a classification of the MPSs of $D=2$ 
with respect to their behavior of two-point correlations.

{\it Theorem.}
Suppose $\mathcal{A}$ is a stochastic unitary (hence CPTP) map.
If $\mathcal{A}^{l}$ does not converge to 
$\mathcal{E}$ in the limit of $\l \rightarrow \infty$ (i.e., $\lim _{l \rightarrow \infty} \mathcal{A}^{l} \neq \mathcal{E}$),
the Kraus operators of $\mathcal{A}$ can always be described 
by choosing a proper basis
as $A[i]/\sqrt{c_i}= X^{ (1- u_3)/2} Z(\phi _i)$, where $u_3=\pm1$ and 
$Z(\phi _i) = e^{ -i \phi _i Z/2}$ with a certain angle $\phi_i$.

{\it Proof.}
A stochastic unitary CPTP map $\mathcal{A}$ can be written in terms of the $\{\frac{1}{2}\sigma_i\}_{i=0}^3$ basis as 
a $4 \times 4$ matrix,
\begin{eqnarray*}
\mathcal{A} = 
\left( 
\begin{array}{cccc}
1 & 0 & 0 & 0 
\\
0 & & &
\\
0 & & \tilde{\mathcal{A}} &
\\
0 & & &
\end{array}
\right),
\end{eqnarray*}
where $\tilde{\mathcal{A}}$ is a $3 \times 3$ real matrix with 
$||\tilde{\mathcal{A}}||_\infty \leq 1$,
where
\begin{eqnarray*}
\|O\|_\infty\equiv\max_{v}\frac{\|Ov\|}{\|v\|}
\end{eqnarray*}
is the $\infty$ norm
(see Appendix C for the matrix representation of the map $\mathcal{A}$).

If $||\tilde{\mathcal{A}}||_\infty <1$, $\mathcal{A}^{l} \rightarrow \mathcal{E}$
in the limit of $l \rightarrow \infty$,
since 
\begin{eqnarray*}
|| \tilde{\mathcal{A}} ^{l} ||_\infty \leq || \tilde{\mathcal{A}}||_\infty^l \rightarrow 0.
\end{eqnarray*}
Hence, in this case, the corresponding MPS state is at least asymptotically normalizable.

On the other hand if $||\tilde{\mathcal{A}}||_\infty =1$,
there exist two vectors $v$ and $v'$ such that $\tilde{\mathcal{A}} v=v'$ 
and $||v||^2 = ||v'||^2 \equiv \bar v$.
We can always choose the vector $v$ and $v'$ as real vectors,
because
if $v$ is a complex vector,
the complex conjugate $v^*$ also subjects to $\tilde {\mathcal{A}} v^* = {v'} ^*$.
Thus we can redefine a real vector $\tilde v$ by $\tilde v \equiv v+v^*$,
which also satisfy $\tilde{\mathcal{A}} \tilde v = \tilde v'$.

Let us consider two Hermitian operators 
\begin{eqnarray*}
V &\equiv&\frac{1}{\sqrt{\bar{v}}}\sum _{i=1}^{3}  v_i \sigma _i ,
\\
V' &\equiv&\frac{1}{\sqrt{\bar{v}}}\sum _{i=1}^{3}  v'_i \sigma _i.
\end{eqnarray*}
By their definitions,
${\rm Tr} [V] ={\rm Tr}[V']=0$ and ${\rm Tr}[V^2]={\rm Tr}[V'^2]=2$,
and therefore
both $\rho _v \equiv  (I+ V)/2$ and $\rho _{v'} \equiv (I + V')/ 2$
are pure states.
Since $\mathcal{A} v=v'$ (i.e., $\mathcal{A} V =V'$), 
$\mathcal{A} \rho _v = \rho _{v '}$;
that is, a pure state $\rho _v$ is mapped 
to another pure state $\rho _{v '}$ without any leakage (trace-preserving).
On the other hand, we have assumed that the Kraus operator $A[i]$ of the map $\mathcal{A}$ is proportional to a unitary 
operator.  
Thus the Kraus operator (MPS matrix) has to be written as 
\begin{eqnarray*}
\frac{1}{\sqrt{c_i}}A[i] = |v' \rangle \langle v| + e^{ i \beta _i} |v'^{\perp}\rangle \langle v ^{\perp} |
\end{eqnarray*}
with a certain angle $\beta _i$ for all $i$,
where $|v \rangle \langle v| \equiv \rho _v$ and 
$|v' \rangle \langle v' | \equiv \rho _{v'}$,
and $|v ^{\perp} \rangle $ and $|{v' }^{\perp}\rangle$ 
are their orthogonals respectively.

By redefining Pauli operators such that $|v\rangle$ and $|v^{\perp}\rangle$
are the eigenstates of $\sigma _3$,
the MPS matrix can be reformulated as 
\begin{eqnarray*}
\frac{1}{\sqrt{c_i}}A[i] = U Z(\beta _i),
\end{eqnarray*}
where $Z(\beta _i ) = e^{ -i \beta _i Z/2}$ and $U=|v' \rangle \langle v|+ |{v'} ^{\perp}  \rangle \langle v ^{\perp}|$.
Then, the map $\mathcal{A}$ can be decomposed into $\mathcal{U} \circ \mathcal{Z}$,
where $\mathcal{Z} = \sum _i Z(\beta _i) \rho Z^{\dag}(\beta _i)c_i$ 
and $\mathcal{U} \rho  =U \rho U^{\dag} $.
The $3\times 3$ matrices $\tilde{\mathcal{Z}}$ and $\tilde{\mathcal{U}}$ are also defined 
for $\mathcal{Z}$ and $\mathcal{U}$, respectively.
Specifically, the $3\times 3$ matrix $\tilde{\mathcal{Z}}$ 
is written in the present basis as
\begin{eqnarray*}
\tilde {\mathcal{Z}}=\left(
\begin{array}{ccc}
 p & -q & 0
\\
 q & p & 0
\\
 0 & 0 & 1
\end{array}
\right) ,
\end{eqnarray*} 
where $p= \sum _i \cos (\beta _i )c_i$
and $q=\sum _i \sin (\beta _i )c_i $.
By redefining the $\sigma _2$ and $\sigma _3$ basis,
the above matrix can be rewritten as
\begin{eqnarray*}
\tilde{\mathcal{Z}}=\left(
\begin{array}{ccc}
 \lambda _1 & 0 & 0
\\
 0 & \lambda _2 & 0
\\
 0 & 0 & 1
\end{array}
\right) .
\end{eqnarray*} 
Note that
\begin{eqnarray*}
1 - p^2 -q^2 &=& \sum _i c_i (\cos ^2 \beta _i + \sin ^2 \beta _i)
- \sum _i c_i (p^2 + q^2)\\
&=&  \sum _i c_i [(\cos \beta _i -p)^2 + (\sin \beta _i -q)^2]\\
&>&0,
\end{eqnarray*}
where
we have used $\cos ^2 \beta _i +\sin^2 \beta _i =1$, $\sum _i c_i =1$,
and $\beta _i \neq \beta _j$.
This means that
\begin{eqnarray*}
|\lambda _1| = |\lambda _2 |=\sqrt{p^2 +q^2}<1.
\end{eqnarray*}
(The reason why we have assumed $\beta_i\neq\beta_j$ is that
if $\beta_1=\beta_2$, for example, then by changing the
physical basis as
\begin{eqnarray*}
|1\rangle&\to&\frac{\sqrt{c_1}}{\sqrt{c_1+c_2}}|\tilde{1}\rangle+\frac{\sqrt{c_2}}{\sqrt{c_1+c_2}}|\tilde{2}\rangle,\\
|2\rangle&\to&\frac{\sqrt{c_2}}{\sqrt{c_1+c_2}}|\tilde{1}\rangle-\frac{\sqrt{c_1}}{\sqrt{c_1+c_2}}|\tilde{2}\rangle,
\end{eqnarray*}
we can make $A[\tilde{2}]=0$
in the MPS.
)

The $3\times3$ matrix $\tilde{\mathcal{U}}$, on the other hand, is an O(3) rotation,
since $\mathcal{U}$ is a unitary operation (i.e. it does not change $|| v ||$).
Let us consider a Hermitian matrix
\begin{eqnarray*}
&&
(\tilde{\mathcal{Z}} \tilde{\mathcal{U}} \tilde{\mathcal{Z}}')^{\dag}
\tilde{\mathcal{Z}} \tilde{\mathcal{U}} \tilde{\mathcal{Z}}'
\\&=& 
\tilde{\mathcal{Z}'}
\tilde{\mathcal{U}}^{T} \tilde{\mathcal{Z}'}^2 \tilde{\mathcal{U}}
\tilde{\mathcal{Z}'}
\\
&=& \bar \lambda \tilde {\mathcal{Z}'}^2 + 
\tilde{\mathcal{Z}}'
\tilde{\mathcal{U}}^{T} 
\left( \begin{array}{ccc}
0 & 0& 0
\\
0  & 0 & 0
\\
0 & 0& 1- \bar \lambda 
\end{array}\right)
\tilde{\mathcal{U}}
\tilde{\mathcal{Z}}'
\\
&=&
\bar \lambda \tilde{\mathcal{Z}'}^2 + (1- \bar \lambda )
\tilde{\mathcal{Z}}'
\left( \begin{array}{ccc}
u_{1}^2 &  u_{2} u_{1}&  u_{3} u_{1}
\\
u_{2} u_{1} & u_{2}^2 &  u_{3}u_{2}
\\
u_{1}u_{3} & u_{3}u_{2} & u_{3}^2
\end{array}\right)
\tilde{\mathcal{Z}}'
\end{eqnarray*}
where $u_{i} = \tilde{\mathcal{U}}_{3i}$, $\bar \lambda = |\lambda _{1,2}|^2$
and $\tilde{\mathcal{Z}}' \equiv {\rm diag}
(\sqrt{\bar \lambda },\sqrt{\bar \lambda },1)$.
The eigenvalues of the above matrix are given by $\bar \lambda ^2$ and
\begin{eqnarray*}
\frac{1}{2} \Bigl[ u_3 ^2(1-\bar \lambda )^2 +
2 \bar\lambda
\pm \left(1-\bar \lambda \right) \sqrt{
(1- \bar \lambda )^2 u _3 ^4 + 4 \bar \lambda u_3 ^2
} \Bigr] \equiv r_\pm.
\end{eqnarray*}
Since 
\begin{eqnarray*}
1=u_1^2+u_2^2+u_3^2\ge u_3^2,
\end{eqnarray*}
\begin{eqnarray*}
|r_\pm|\le\frac{1}{2}\Big[(1-\bar{\lambda})^2+2\bar{\lambda}+
(1-\bar{\lambda})\sqrt{(1-\bar{\lambda})^2+4\bar{\lambda}}\Big]=1.
\end{eqnarray*}
If $|r_+|=1$ or $|r_-|=1$,
$u_3 =\pm 1$, and therefore $u_1=u_2=0$.
This implies that 
\begin{eqnarray*}
U = X^{ (1- u_3)/2} Z(\phi)
\end{eqnarray*}
with a certain $\phi$. 
In this case, $A[i]/\sqrt{c_i}= X^{ (1- u_3)/2} Z(\phi _i)$
with a certain angle $\phi _i$.
Therefore, 
$\mathcal{A}^{l}$ does not converge toward $\mathcal{E}$,
since $\mathcal{A}$ has eigenoperators with eigenvalues $\pm1$.
On the other hand, if $|r_\pm|<1$, $||\tilde{\mathcal{Z}} \tilde{\mathcal{U}} \tilde{\mathcal{Z}}'||<1$.
In this case,
\begin{eqnarray*}
|| \tilde {\mathcal{A}} ^l || &\leq&  
||\tilde{\mathcal{U}}  \tilde{\mathcal{Z}} 
\tilde{\mathcal{U}} \tilde{\mathcal{Z}}||^{\lfloor l/2 \rfloor}
\\
&\leq&
 || \tilde{\mathcal{Z}} 
\tilde{\mathcal{U}} \tilde{\mathcal{Z}}||^{\lfloor l/2 \rfloor}
\\
&\leq& || \tilde{\mathcal{Z}} 
\tilde{\mathcal{U}} \tilde{\mathcal{Z}}'||^{\lfloor l/2 \rfloor} \rightarrow 0,
\end{eqnarray*}
and therefore
$\mathcal{A}^l \rightarrow \mathcal {E}$ in the limit of $l \rightarrow 0$ $\blacksquare$.
	
This theorem shows that if a resource state 
is not normalizable even in the limit of infinite size,
the MPS matrix is always given by
$A[i]/\sqrt{c_i}= X^{ (1- u_3)/2} Z(\phi _i)$.
Apparently, such an MPS whose matrix is
$A[i]/\sqrt{c_i}= X^{ (1- u_3)/2} Z(\phi _i)$
cannot allow an arbitrary single-qubit rotation on it.
In other words, all resource states
which are capable of an arbitrary single-qubit rotation are 
at least asymptotically normalizable (i.e. $\displaystyle\lim _{l \rightarrow \infty} \mathcal{A}^l = \mathcal{E}$).
That is, normalizability,
which is a sufficient condition for correct projective measurements in the CSs, 
in the limit of $l \rightarrow \infty$ 
is also a necessary condition for an arbitrary single-qubit rotation.
Accordingly, normalizability (at least asymptotic sense) is a necessary and sufficient
condition for a universal quantum computational wire,
where one can simulate correct projective measurements
and arbitrary single qubit rotations.
In addition, the theorem combined with
the relationship between
the map $\mathcal{A}$ and two-point correlations
(see Appendix B)
tells us another
interesting fact.
When $\lim _{l \rightarrow } \mathcal{A} ^{l} = \mathcal{E}$
two-point correlations always decay exponentially (see Appendix D for details).
On the other hand, if $\lim _{l \rightarrow } \mathcal{A} ^{l} \neq \mathcal{E}$,
 the Kraus operator of the map $\mathcal{A}$ 
is always given by $A[i]/\sqrt{c_i}= X^{ (1- u_3)/2} Z(\phi _i)$,
and hence the two-point correlations do not decay.
Since any stochastic unitary map $\mathcal{A}$
can be classified into these two cases due to the theorem,
no MPS of polynomially decaying two-point correlations
can be described by $\mathcal{A}$ of a stochastic unitary map.
Thus such an MPS cannot be used for MBQC in this framework. 

The above results highlight an interesting relationship 
between computational power and correlation in resource states:
the many-body states 
whose two-point correlations do not decay 
such as the GHZ and the Ising antiferromagnetic states
cannot be a universal resource for QCTN.
Actually, the MPS matrices of the GHZ state are given by $A[0]=I/\sqrt{2}$ and $A[1]=Z/\sqrt{2}$,
which apparently does not allow an arbitrary single-qubit rotation in CSs.
In addition, a many-body state
whose two-point correlations exhibit polynomial decay,
such as the $W$ state,
does not allow unitary operations in the CS.
Accordingly, from the theorem, we can conclude that a many-body state 
whose two-point correlations cannot be upperbounded by an exponentially decaying
function, such as the GHZ, the Ising antiferromagnetic, and the $W$ states,
cannot be useful for QCTN,
which is the most general framework for MBQC to date \cite{comment_QCTN}.
This also suggests that a ground state near the critical point of a quantum phase transition
cannot be used as a resource state for MBQC, since two-point correlations of such a state decay as a polynomial function \cite{QPT}. 
Evidence that computational power in
MBQC is a robust property of a quantum phase has
been demonstrated recently on the Haldane phase \cite{Bartlett}.
Since a two-point correlation is a good witness of quantum phase transitions,
the present result suggests that such a robust property should
hold in other general models of quantum phase transitions.

\section{Conclusion}
We have investigated relationships between computational power
and two-point correlations in the resource states for QCTN
considering possibilities of projective measurements 
and arbitrary single-qubit 
rotations, both of which are important properties of a universal resource.
Our findings are as follows: 
(i) We have shown that the normalizable resources with a finite $l$
exhibit exactly vanishing two-point correlations
with a finite distance like the cluster state,
and always allow correct projective measurements 
in the CS although they are of finite size.
(ii) We have explored how to perform projective measurements
on the finite-size resource states with non vanishing two-point correlations
by introducing a class of MPSs.
(iii) We have shown that all resource states which are capable of an arbitrary single-qubit rotation
are at least asymptotically normalizable and exhibit 
exponentially decaying two-point correlations.
Accordingly, normalizability (at least asymptotic sense) is a necessary and sufficient condition
for a universal quantum computational wire which allows simulations of both
correct projective measurements and arbitrary single-qubit rotations.
(iv) Further, we have seen that a many-body state whose two-point correlations cannot
be upperbounded by an exponentially decaying function
does not allow an arbitrary single-qubit rotation,
since the matrices of such an MPS cannot be described 
as unitary operators.
These results (iii) and (iv) have been obtained
by showing the theorem about convergence of a stochastic unitary map.
This theorem itself seems to have an interesting meaning in the context of quantum channel.
The detailed study is, however, out of our scope and an interesting topic for future work.

Let us finally mention the
extension of the above results to tensor network states.
Since simultaneous single-qubit measurements are enough
to readout results of quantum computation,
we can apply the present argument for 
tensor network states by reducing them to MPSs before the readout.
Thus similar results would be obtained also for tensor network states.

We believe that the present results give us a clue
to find a novel resource state for QCTN and also help us
to build a bridge between quantum information science
and many-body physics, such as the physics of quantum phase transitions.

\begin{acknowledgments}
KF is supported by MEXT Grant-in-Aid for Scientific Research on Innovative Areas 20104003.
TM is supported by ANR (StatQuant JC07 07205763)
and JSPS.
\end{acknowledgments}

\appendix

\section{Normalizability and completely depolarizing channel}
If an MPS is normalizable,
\begin{eqnarray*}
\frac{1}{2}
=
f_{n-k}(|R\rangle , |\psi \rangle )=
{\rm Tr} \Big[ |R\rangle \langle R| \mathcal {A}^{\circ (n-k)} |\psi \rangle \langle \psi | \Big]
\end{eqnarray*}
for all $|R\rangle$ and $|\psi \rangle$.
This means that 
\begin{eqnarray*}
\mathcal A^{\circ (n-k)} \rho = \frac{I}{2} {\rm Tr} [\rho]
\end{eqnarray*}
for all $\rho$.
Then $\mathcal A^{\circ (n-k)}$ is a completely depolarizing channel, because
\begin{eqnarray*}
\mathcal {A}^{\circ (n-k)} \rho &=& 
\frac{I}{2}{\rm Tr} [\rho]
\\
&=& \sum _{i} \langle i | V  \rho  V^{\dag}| i \rangle  U\frac{I}{2}U^{\dag}
\\
&=& \frac{1}{2}\sum _{ij} U|j\rangle \langle i | V \rho V^{\dag} |i \rangle \langle j | U^{\dag}
\\
&=&
\frac{1}{4} \sum _{i=0}^{3}
U \sigma _{i} V \rho V^{\dag} \sigma _{i} U^{\dag} .
\end{eqnarray*}

\section{Normalizability and two-point correlation}
The expectation value of an observable $O^{a}_k O^{b} _{k+r}$
with respect to $| \bar \Psi (R ,L) \rangle _1 ^n$ can be calculated as
\begin{eqnarray*}
&&
\langle O^{a}_k O^{b}_{k+r} \rangle _{\Psi} 
\\
&\equiv&
{\rm Tr} \Big[ O^{a}_k O^{b} _{k+r} |  \bar \Psi (R,L) \rangle _1^{n} \langle  \bar \Psi (R,L) | \Big] 
\\
&=&
\frac{1}{f_n(|R\rangle,|L\rangle)}
\langle R |\Big[
\mathcal{A}^{n-(k+r)}\mathcal{O}^{b}_{k+r} \mathcal{A}^{r-1} \mathcal{O}^{a}_{k} \mathcal{A}^{k-1}|L\rangle \langle L|\Big] |R\rangle ,
\end{eqnarray*}
where 
$O^{a}_{k}$ and $O^{b}_{k+r}$ are arbitrary
operators on the $k$th and $(k+r)$th qudits respectively, 
and
\begin{eqnarray*}
\mathcal{O}_{k}^{a,b} \rho \equiv \sum _{ij} 
\langle   i |O_k^{a,b} |j\rangle  A[j] \rho A^{\dag}[i] .
\end{eqnarray*}
Since $n>l$, the normalization factor becomes 
\begin{eqnarray*}
f_n(|R\rangle , |L\rangle)=\frac{1}{2}.
\end{eqnarray*}
When $r >l$,
\begin{eqnarray*}
\mathcal{A}^{r-1} =A^{r-1-l} \mathcal{E} 
\end{eqnarray*}
for a normalizable resource with a finite $l$.
By using this, we obtain
\begin{eqnarray*}
\langle O^{a}_k O^{b}_{k+r} \rangle  _{\Psi}
&=&
2 \langle R| \Big[
 \mathcal{A}^{n-(k+r)}    
\mathcal{O}^b_{k+r} \frac{I}{2}  \Big]|R\rangle
\\
&&\times
{\rm Tr} \Big[\mathcal{O}^{a}_{k} \mathcal{A}^{k-1} |L\rangle \langle L|\Big].
\end{eqnarray*}
On the other hand, the expectation value of $O^a_k$ and $O^b_{k+r}$
can be obtained respectively as
\begin{eqnarray*}
\langle O^{a}_k \rangle  _{\Psi}& \equiv &
{\rm Tr} \Big[ O^{a}_k  | \bar \Psi (R,L) \rangle _1^{n} \langle  \bar \Psi (R,L) | \Big] 
\\
&=& 
{\rm Tr} \Big[\mathcal{O}^a_{k} \mathcal{A}^{k-1} |L\rangle \langle L|\Big],
\end{eqnarray*}
and
\begin{eqnarray*}
\langle O^b _{k+r} \rangle  _{\Psi}
&=& 
{\rm Tr} \Big[ O^{b}_{k+r}  | \bar \Psi (R,L) \rangle _1^{n} \langle \bar \Psi (R,L) | \Big] 
\\
&=& 2 \langle R| \Big[
\mathcal{A}^{n-(k+r)}    
\mathcal{O}_{k+r}^{b}\frac{I}{2}\Big] |R\rangle.
\end{eqnarray*}
As a result,
\begin{eqnarray*}
\langle O^a_k O^b_{k+r} \rangle _{\Psi} - \langle O^a_k \rangle _{\Psi} \langle O^b_{k+r} \rangle _{\Psi}=0
\end{eqnarray*}
for all $r >l$.

\section{Matrix representation of CPTP map}
Consider a unital CPTP map $\mathcal{A} \rho = \sum _{i} A[i] \rho A^{\dag} [i]$,
where $\sum _i A^{\dag} [i]A[i]=I$ and $A[i]A^{\dag}[i] \propto I$.
Each Kraus operator can be decomposed into
the Pauli matrices as
\begin{eqnarray}
A[i] = \sum _j \alpha _{ij} \sigma _j.
\end{eqnarray}
The CPTP condition reads 
\begin{eqnarray}
&&\sum _{ij} \alpha _{ij} \alpha^* _{ij} =1,
\label{Cond1}
\\
&&\sum _{ijk} \alpha _{ij} \alpha ^*_{ik} \epsilon _{jkl}  +\alpha _{i0} \alpha ^* _{il} +  \alpha  _{il} \alpha _{i0}^*= 0.
\label{Cond2}
\end{eqnarray}
The action of $A[i]$ under its conjugation is given by
\begin{eqnarray*}
A[i] \sigma _k A^{\dag}[i] = \sum _{jl} \alpha _{ij}\alpha ^*_{il}
 \sigma _j \sigma _k \sigma _l .
\end{eqnarray*}
By using this and  condition (\ref{Cond1}),
we obtain
$\mathcal{A}_{00} = \sum _{ij} \alpha _{ij}  \alpha _{ij}^*=1$.
Furthermore, by using condition (\ref{Cond2}),
we obtain
$\mathcal{A}_{0k} = \sum _i
\alpha _{ik}\alpha ^*_{i0} + \alpha _{i0}\alpha ^*_{ik}
+ \alpha _{ij}\alpha ^*_{il} \epsilon _{jlk}=0$.
Similarly we can also obtain $\mathcal{A}_{k0} =0$.

Next we define
a $3\times3$ matrix $\tilde{\mathcal{A}}$ by 
$\mathcal{A}_{ij}$ for $i,j\neq 0$.
A quantum state $\rho$ (i.e. an hermitian operator with trace one) 
is represented 
in terms of the Pauli basis as $\rho = I/2 + \sum v_i \sigma _i /2$,
where $v_i$ is real for all $i$ and its purity $\leq 1$.
Since the action of $A[i]$ map an hermitian operator to another hermitian operator
under its conjugation ,
$\tilde{\mathcal{A}}_{ji} v_i$ is also real for all $i$ and $j$.
It requires that $\tilde {\mathcal{A}}$ is a real matrix.
Since a CPTP map $\mathcal{A}$ transforms a state $\rho$
to another physical state $\mathcal{A} \rho $ without increasing its purity,
$|| \tilde{\mathcal{A}} v|| /|| v|| \leq 1$ for all $v$,
which implies that $||\tilde{\mathcal{A}}|| \leq 1$.

\section{Correlation}
Let us consider the correlation
\begin{eqnarray*}
\langle O_k^aO_{k+r}^b\rangle_\Psi
-\langle O_k^a\rangle_\Psi\langle O_{k+r}^b\rangle_\Psi
\end{eqnarray*}
between two Hermitian operators $O_k^a$ on site $k$ and
$O_{k+r}^b$ on site $k+r$ ($\|O_k^a\|_\infty\le1$, $\|O_{k+r}^b\|_\infty\le1$),
where 
\begin{eqnarray*}
\langle O\rangle_\Psi
=
\langle\bar{\Psi}(R,L)|O|\bar{\Psi}(R,L)\rangle_1^n.
\end{eqnarray*}
Let us define the (super)operator
\begin{eqnarray*}
{\mathcal O}\rho=\sum_{i=0}^{d-1}\sum_{j=0}^{d-1}
\langle i|O|j\rangle A[j]\rho A^\dagger[i].
\end{eqnarray*}
Then
\begin{eqnarray*}
\langle O_k^aO_{k+r}^b\rangle_\Psi=
\frac{1}{f_n(|R\rangle,|L\rangle)}
\langle R|\Big[{\mathcal A}^{n-(k+r)}{\mathcal O}_{k+r}^b
{\mathcal A}^{r-1}
\\
{\mathcal O}_k^a{\mathcal A}^{k-1}
|L\rangle
\langle L|
\Big]|R\rangle.
\end{eqnarray*}

Let us write
\begin{eqnarray*}
{\mathcal O}_k^a{\mathcal A}^{k-1}|L\rangle\langle L|
=\alpha(O_k^a)\frac{I}{2}+E(O_k^a),
\end{eqnarray*}
where $\alpha(O_k^a)$ is a real number and $E(O_k^a)$
is a trace-less Hermitian operator.
Then,
\begin{eqnarray*}
{\mathcal A}^{r-1}
{\mathcal O}_k^a{\mathcal A}^{k-1}|L\rangle\langle L|
=
\alpha(O_k^a)\frac{I}{2}+
{\mathcal A}^{r-1}E(O_k^a).
\end{eqnarray*}
Note that
\begin{eqnarray*}
\|E(O_k^a)\|_\infty&\le&\|{\mathcal O}_k^a{\mathcal A}^{k-1}|L\rangle\langle L|\|_\infty+\frac{1}{2}|\alpha(O_k^a)|\\
&\le&\sum_{ij}|\langle i|O_k^a|j\rangle|\cdot\|A[j]\rho A^\dagger[i]\|_\infty+\frac{1}{2}|\alpha(O_k^a)|\\
&\le&d\sum_{ij}\sqrt{c_i}\sqrt{c_j}+\frac{1}{2}|\alpha(O_k^a)|\\
&\le&d^2+\frac{1}{2}|\alpha(O_k^a)|,
\end{eqnarray*}
where $\rho$ is a state and we have used
\begin{eqnarray*}
|\langle i|O_k^a|j\rangle|&=&
\Big|\sum_k o_k \langle i|o_k\rangle\langle o_k|j\rangle\Big|\\
&\le&
\sum_k|o_k|\cdot|\langle i|o_k\rangle|\cdot|\langle o_k|j\rangle|\\
&\le&
1~\sum_k\\
&=&d.
\end{eqnarray*}
(here, $o_k$ and $|o_k\rangle$ are eigenvalues and eigenvectors
of $O_k^a$)
and
\begin{eqnarray*}
\sum_{ij}\sqrt{c_i}\sqrt{c_j}\le d.
\end{eqnarray*}
This is because
\begin{eqnarray*}
\sum_ic_i\Big(\frac{1}{\sqrt{c_i}}\Big)^2
-\Big[\sum_ic_i\Big(\frac{1}{\sqrt{c_i}}\Big)\Big]^2\ge0.
\end{eqnarray*}
Thus
\begin{eqnarray*}
&&\langle O_k^aO_{k+r}^b\rangle_\Psi
\\
&=&
\frac{\alpha(O_k^a)}{2f_n(|R\rangle,|L\rangle)}
\langle R|\Big[{\mathcal A}^{n-(k+r)}{\mathcal O}_{k+r}^b
I\Big]|R\rangle
\\
&&+
\frac{1}{f_n(|R\rangle,|L\rangle)}
\langle R|\Big[{\mathcal A}^{n-(k+r)}{\mathcal O}_{k+r}^b
{\mathcal A}^{r-1}E(O_k^a)\Big]|R\rangle\\
&\equiv&\frac{C_1}{2f_n(|R\rangle,|L\rangle)}+\frac{C_2}{f_n(|R\rangle,|L\rangle)}.
\end{eqnarray*}
From this, we also obtain
\begin{eqnarray*}
\langle O_k^a\rangle_\Psi
\\
&=&
\frac{\alpha(O_k^a)}{2f_n(|R\rangle,|L\rangle)}
\\
&&+
\frac{1}{f_n(|R\rangle,|L\rangle)}
\langle R|\Big[{\mathcal A}^{n-k}E(O_k^a)
\Big]|R\rangle\\
&\equiv&\frac{\alpha(O_k^a)}{2f_n(|R\rangle,|L\rangle)}
+\frac{C_3}{f_n(|R\rangle,|L\rangle)}
\end{eqnarray*}
and
\begin{eqnarray*}
&&\langle O_{k+r}^b\rangle_\Psi
\\
&=&
\frac{\alpha(I)}{2f_n(|R\rangle,|L\rangle)}\langle R|\Big[{\mathcal A}^{n-(k+r)}{\mathcal O}_{k+r}^b
I\Big]|R\rangle
\\
&& +
\frac{1}{f_n(|R\rangle,|L\rangle)}
\langle R|\Big[{\mathcal A}^{n-(k+r)}{\mathcal O}_{k+r}^b
{\mathcal A}^{r-1}E(I)\Big]|R\rangle\\
&\equiv&\frac{C_4}{2f_n(|R\rangle,|L\rangle)}
+\frac{C_5}{f_n(|R\rangle,|L\rangle)}.
\end{eqnarray*}

Note that
\begin{eqnarray*}
f_n(|R\rangle,|L\rangle)&=&\langle R|({\mathcal A}^n|L\rangle\langle L|)|R\rangle\\
&=&\langle R|\Big(\frac{I}{2}+{\mathcal A}^nF\Big)|R\rangle\\
&=&\frac{1}{2}+\langle R|{\mathcal A}^nF|R\rangle,
\end{eqnarray*}
where $F$ is a trace-less Hermitian operator.
Therefore,
\begin{eqnarray*}
\Big|f_n(|R\rangle,|L\rangle)-\frac{1}{2}\Big|&=&
\Big|\langle R|{\mathcal A}^nF|R\rangle\Big|\\
&\le&
\|{\mathcal A}^nF\|_\infty\\
&\le&
\frac{1}{2}\|(\tilde{{\mathcal A}})^n\|_\infty,
\end{eqnarray*}
where $\tilde{\mathcal A}$ is the $3\times 3$ matrix defined in the previous
section.

Furthermore,
\begin{eqnarray*}
|\alpha(O_k^a)|&=&
\Big|\mbox{Tr}({\mathcal O}_k^a{\mathcal A}^{k-1}|L\rangle\langle L|)\Big|\\
&=&\Big|\mbox{Tr}\Big(\sum_{i,j}\langle i|O_k^a|j\rangle
A[j]\xi A^\dagger[i]\Big)\Big|\\
&\le&\sum_{i,j}|\langle i|O_k^a|j\rangle|\cdot
\Big|\mbox{Tr}(A[j]\xi A^\dagger[i]\big)\Big|\\
&\le&d\sum_{i,j}\sqrt{c_i}\sqrt{c_j}\\
&\le&d^2,
\end{eqnarray*}
where $\xi$ is a density operator.
And
\begin{eqnarray*}
&&
\Big|\langle R|\Big[
{\mathcal A}^{n-(k+r)}{\mathcal O}_{k+r}^bI
\Big]
|R\rangle
\Big|
\\
&=&
\Big|
\mbox{Tr}\Big(|R\rangle\langle R|{\mathcal A}^{n-(k+r)}{\mathcal O}_{k+r}^b
I
\Big)\Big|\\
&=&
\Big|
\mbox{Tr}\Big(\eta{\mathcal O}_{k+r}^bI\Big)
\Big|\\
&\le&
\sum_{i,j}
|\langle i|O_{k+r}^b|j\rangle|
\cdot
\Big|
\mbox{Tr}\Big(A^\dagger[i]\eta
A[j]\Big)
\Big|\\
&\le&
d
\sum_{i,j}\sqrt{c_i}\sqrt{c_j}\\
&\le&d^2,
\end{eqnarray*}
where $\eta$ is a density operator.
Therefore,
\begin{eqnarray*}
&&
\Big|\frac{C_1}{2f_n(|R\rangle,|L\rangle)}
-
\frac{\alpha(O_k^a)}{2f_n(|R\rangle,|L\rangle)}
\frac{C_4}{2f_n(|R\rangle,|L\rangle)}
\Big|
\\
&=&\frac{|C_1|}{2f_n^2(|R\rangle,|L\rangle)}
\Big|f_n(|R\rangle,|L\rangle)-\frac{1}{2}\Big|\\
&\le&
\frac{d^4}{2(\frac{1}{2}-\frac{1}{2}\|(\tilde{\mathcal A})^n\|_\infty)^2}
\frac{1}{2}\|(\tilde{\mathcal A})^n\|_\infty.
\end{eqnarray*}
In the similar way,
\begin{eqnarray*}
|C_2|
&=&\Big|
\langle R|\Big[
{\mathcal A}^{n-(k+r)}
{\mathcal O}_{k+r}^b
{\mathcal A}^{r-1}
E(O_k^a)
\Big]|R\rangle
\Big|\\
&=&
\Big|
\mbox{Tr}\Big(
\eta{\mathcal O}_{k+r}^b{\mathcal A}^{r-1}E(O_k^a)
\Big)
\Big|\\
&\le&2d^2\|{\mathcal A}^{r-1}E(O_k^a)\|_\infty\\
&\le&\frac{3}{2}d^4\|(\tilde{{\mathcal A}})^{r-1}\|_\infty,
\end{eqnarray*}
where $\eta$ is a state.

\begin{eqnarray*}
|C_3|
&=&\Big|
\langle R|\Big[
{\mathcal A}^{n-k}
E(O_k^a)
\Big]
|R\rangle
\Big|\\
&\le&
\|{\mathcal A^{n-k}}E(O_k^a)\|_\infty\\.
&\le&
\frac{3d^2}{2}\|({\tilde{\mathcal A})^{n-k}}\|_\infty.
\end{eqnarray*}

In conclusion, the two-point correlation 
is upper-bounded by an exponentially decaying function.


\begin{thebibliography}{9}
\bibitem{predict}
In Ref.~\cite{Gross07_1,Gross07_2,Gross07_3},
a universal resource for MBQC is defined that one can efficiently 
predict the result of any quantum computation
with local measurements on such a resource state.
However, the possibility of projective measurements in the CS,
which is one of the important properties for a universal resource,
relies on the arguments for the one-way model in Ref. \cite{Gross07_1,Gross07_2,Gross07_3},
and cannot be applied straightforwardly to general resources.
\bibitem{comment2}
The nondeterminism in the measurement
is treated as usual \cite{Gross07_1,Gross07_2,Gross07_3}.
\bibitem{comment}
In the case of AKLT states,
whether we can perform projective measurements or not
is determined by how we choose the boundary condition.
\bibitem{comment4}
It is interesting to note that for normalizable resources we do not need
to choose a specific boundary vector $|R\rangle$, 
since the two-point correlations
exactly vanish within a finite distance, and therefore the boundary vector
causes no effect. On the other hand, if we consider resource states
with non-vanishing two-point correlations, we have to choose an appropriate boundary vector like this example.
\bibitem{comment_QCTN}
Here we consider computational power of MBQC in QCTN,
which is the most general framework for MBQC to date.
Thus these statements might not hold for other MBQC models,
which might be found in the future.
It is, however, an interesting open question whether or not these statements
based on QCTN can hold at a more general level of MBQC.
\end{thebibliography}
\end{document}